\documentclass[utf8]{beilstein}

\usepackage{xspace}
\usepackage{graphicx}
\usepackage{amsmath}
\usepackage{amssymb}
\usepackage{float}
\usepackage{graphicx}  
\usepackage{capt-of}   
\usepackage{caption}

\begin{document}

\title{Characterization of ion track etched conical nanopores in thermal and PECVD SiO$_{2}$ using small angle X-ray scattering}

\author*[1]{Shankar Dutt}{shankar.dutt@anu.edu.au}

\author[1]{Rudradeep Chakraborty}

\author{Christian Notthoff}
\affiliation{Department of Materials Physics, Research School of Physics, The Australian National University, Canberra ACT 2601, Australia}
\author{Pablo Mota-Santiago}
\affiliation{ANSTO-Australian Synchrotron, Clayton VIC 3168, Australia}
\author{Christina Trautmann}
\affiliation{GSI Helmholtzzentrum für Schwerionenforschung, Planckstr. 1, 64291, Darmstadt, Germany}
\affiliation{Technische Universtät Darmstadt, 64289, Darmtadt, Germany}
\author[1]{Patrick Kluth}

\maketitle

\begin{abstract}
Conical nanopores in amorphous SiO$_{2}$ thin films fabricated using the ion track etching technique show promising potential for filtration, sensing and nanofluidic applications. The characterization of the pore morphology and size distribution, along with its dependence on the material properties and fabrication parameters, is crucial to designing nanopore systems for specific applications. Here, we present a comprehensive study of track-etched nanopores in thermal and plasma-enhanced chemical vapor deposited (PECVD) SiO$_{2}$ using synchrotron-based small-angle X-ray scattering (SAXS).  We present a new approach for analyzing the complex highly anisotropic 2-dimensional SAXS patterns of the pores by reducing the analysis to two orthogonal 1-dimensional slices of the data. The simultaneous fit of the data enables an accurate determination of the pore geometry and size distribution. The analysis reveals substantial differences between the nanopores in thermal and PECVD SiO$_{2}$. The track-to-bulk etching rate ratio is significantly different for the two materials, producing nanopores with cone angles that differ by almost a factor of 2. Furthermore, thermal SiO$_2$ exhibits an exceptionally narrow size distribution of only $\sim$2-4\%, while PECVD SiO$_2$ shows a higher variation ranging from $\sim$8-18\%. The impact of ion energy between 89 MeV and 1.6 GeV on the size of the nanopores was also investigated for pores in PECVD SiO$_{2}$ and shows only negligible influence. These findings provide crucial insights for the controlled fabrication of conical nanopores in different materials, which is essential for optimizing membrane performance in applications that require precise pore geometry.
\end{abstract}


\section{Introduction}
Solid-state nanopores have attracted significant attention in the past decade due to their broad applicability in a variety of areas including biosensing, micro/ultrafiltration, desalination, ion and molecular separation, dialysis, battery technologies, blue energy generation, and nanofluidics \cite{xue_solid-state_2020,he_solid-state_2021,yeh_electro-osmotic_2015,zhao_nanopore_2024,wanjiya_nanofiltration_2024,dutt_shape_2021,dutt_ultrathin_2023,wei_engineering_2023,meyer_solid-state_2021,siwy_nanopores_2023,varongchayakul_single-molecule_2018,lu_recent_2022,dutt_annealing_2023,pan_osmotic_2024,zhang_fundamental_2016,mohammad_nanofiltration_2015,calvo_comparison_2004,kanani_permeabilityselectivity_2010,dutt_high_2023,li_conical_2004,wang_saxs_2022,sun_selective_2019,desormeaux_nanoporous_2014,wang_nanopore-based_2021,wang_biosensor_2022,kiy_ion_2021,rastgar_harvesting_2023,tsutsui_discriminating_2017,hadley_analysis_2020,kiy_highly_2023,liu_dynamic_2024,vlassiouk_versatile_2009,dutt_nanopore_2023}. Conical nanopores are of particular interest due to the asymmetric ion transport resulting from their unique geometry.  \cite{kiy_highly_2023,ma_single_2010,balannec_nanofiltration_2018,zhang_conical_2017,lin_rectification_2019,duleba_effect_2022}. 

Conical nanopores can be reproducibly fabricated at scale using the track-etch technology in a number of different materials \cite{hellborg_ion_2010,hadley_analysis_2020}. This method involves irradiating the material with swift heavy ions to create long and narrow damaged regions along the paths of the ions known as 'ion tracks'. These ion tracks are more susceptible to chemical etching compared to the undamaged material, which can be exploited for the fabrication of nanopores with narrow size distribution \cite{dutt_annealing_2023, hadley_analysis_2020, hadley_etched_2019}. The geometry of the resulting nanopores is determined by several factors, including the substrate material, the type and concentration of the etchant, the density of the material and the type and energy of the ions used \cite{dutt_annealing_2023,hadley_analysis_2020,hadley_etched_2019}. 

Track-etch technology has been used for the commercial fabrication of cylindrical nanopores in polymers for filtration applications \cite{apel_micro-_2011,apel_swift_2003,liu_fabrication_2019,hanot_industrial_2009,kaya_reviewtrack-etched_2020,price_recent_2008}. Only recently we have adapted this technology to generate conical nanopores in silicon dioxide \cite{hadley_analysis_2020,hadley_etched_2019,kiy_highly_2023}. Amorphous silicon dioxide (SiO$_{2}$)  has excellent chemical stability, well-understood surface chemistry, and compatibility with semiconductor processing, opening up new applications for track-etched nanopores in this material \cite{kiy_highly_2023}. 

In this study, we report the characterization of track-etched nanopores in two types of silicon dioxide: one produced by wet thermal oxidation of Si (thermal SiO$_2$) and another deposited by Plasma-Enhanced Chemical Vapor Deposition (PECVD). Thermally grown SiO$_2$ is of high quality and stoichiometric, however, requires high temperatures for growth, and can only be grown on a Si substrate. PECVD, on the other hand, allows deposition at much lower temperatures on many different substrates with control over the film properties, such as stoichiometry, density, refractive index, and residual stress. As these fabrication methods involve fundamentally different growth mechanisms, the resulting layers have different properties \cite{mota-santiago_nanoscale_2018,karouta_structural_2012} and it can be expected that the track etched nanopores also show different characteristics, including the track etching process itself. Understanding how the different fabrication methods influence the characteristics of ion-track etched nanopores is crucial to optimize their fabrication for specific applications. Here we focus on characterizing the size, geometry and size distribution of track etched nanopores in thermal and PECVD SiO$_2$ as these parameters are critical for membrane performance in specific applications, including selectivity, throughput and molecular capture. 

Small-angle X-ray scattering (SAXS) has proven to be an invaluable tool for characterizing nanopore membranes, offering nondestructive analytical capabilities that yield statistical information of >$10^7$ pores \cite{dutt_shape_2021,dutt_annealing_2023,hadley_analysis_2020,hadley_etched_2019}. Our previous work demonstrated the effectiveness of SAXS for studying conical nanopores in SiO$_2$, providing unprecedented precision in determining the pore morphologies \cite{hadley_analysis_2020,hadley_etched_2019}. The method involved fitting two-dimensional scattering patterns to a conical pore model utilizing a series of images with different tilts of the sample with respect to the incident X-ray beam, corresponding to the alignment of the parallel pores with the beam. Although highly accurate, this approach has two limitations: the computational resources required for numerical calculation of intensity values for each pixel in the 2D fit and the challenge of incorporating size distribution analysis due to the computational complexity of applying distribution functions in a 2D fitting scenario. To address these limitations, we have developed a new approach that maintains the high precision of SAXS analysis while significantly reducing computational requirements and enabling the investigation of size distributions. This method involves analyzing and fitting 1D sections of the SAXS patterns employing different form factors rather than performing 2D image fitting.

We implemented our new fitting method to investigate conical nanopores in the two different SiO$_2$ membrane materials. For nanopores in thermal SiO$_2$ we confirm that these results are consistent with our previous studies that employ 2D fitting and quantify the size distribution. Track-etched nanopores in PECVD-SiO$_2$ have not been studied before and revealed striking differences in the geometrical parameters due to a different track-to-bulk etching rate ratio and a wider size distribution.

\section{Results and Discussions}

\begin{figure}
    \centering
    \includegraphics[width=\textwidth]{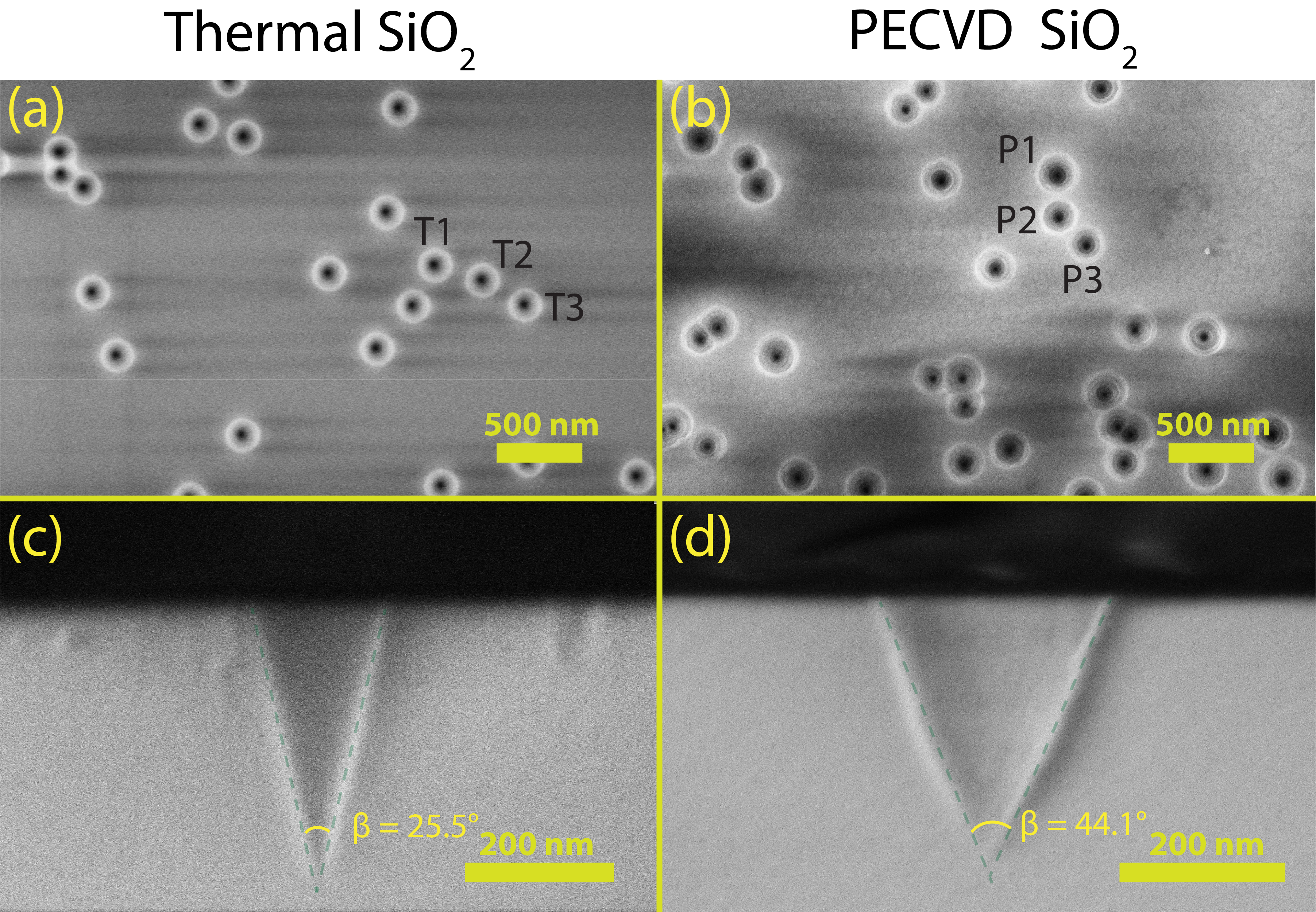}
    \caption{Plan‐view (a, b) and cross‐sectional (c, d) scanning electron microscopy images of nanopores in thermal (a, c) and PECVD (b, d) SiO$_2$. The thermal and PECVD SiO$_{2}$ thin films were irradiated with 1.6 GeV Au ions and subsequently etched in 3\% HF for 8.5 min and 6 min respectively. The top‐view images (a, b) highlight the circular pore openings, while the cross‐sectional views (c, d) reveal conical pore geometry. The cone‐angle ($\beta$) of the conical pores in thermal SiO$_2$ (c) is approximately 1.8 times less than that in PECVD SiO$_2$ (d).}
    \label{fig:1DSAXS1}
\end{figure}

Figures \ref{fig:1DSAXS1} (a, b) show plan view scanning electron microscopy (SEM) images of nanopores fabricated in thermal (a) and PECVD (b) SiO$_{2}$ by etching ion tracks produced with 1.6 GeV Au ion irradiation (see Experimental and Theory section for details). Although both materials reveal conical nanopores, thermal SiO$_{2}$ exhibits more uniformly sized nanopores, while PECVD SiO$_{2}$ displays higher dispersity in pore dimensions (excluding overlapping pores). As an example, three representative non-overlapping pores are highlighted in Figures \ref{fig:1DSAXS1}(a) and (b). Their average radius at the sample surface—T1, T2, and T3—are measured to be 100.3 $\pm$ 1.3 nm, 103.2 $\pm$ 1.6 nm, and 102.4 $\pm$ 2.6 nm, respectively, in thermal SiO$_{2}$. In contrast, the representative pores (P1, P2 and P3) in PECVD SiO$_{2}$  measure 127.2 $\pm$ 2.7 nm, 112.3 $\pm$ 2.4 nm, and 114.7 $\pm$ 1.6 nm. Although only three pores are shown, they illustrate the larger size variation in PECVD SiO$_{2}$  compared to the uniform pore size in thermal SiO$_{2}$. From SEM measurements, the standard deviation in the pore radius was measured to be $\sim$1.8 nm for thermal SiO$_{2}$ but $\sim$8 nm for PECVD SiO$_{2}$. The reader must note that, unless otherwise noted, the nanopore radius or size mentioned throughout this work refers specifically to the radius of the cone base. To overcome the limited sampling of pores in SEM imaging, we complemented the microscopy analysis with small-angle X-ray scattering, which provides statistically robust measurements, averaging over more than 10\textsuperscript{7} pores during an experiment.

Cross-sectional SEM images (Fig. \ref{fig:1DSAXS1}(c,d)) reveal distinct differences in nanopore geometry between thermal and PECVD SiO2$_{2}$. The full cone angle ($\beta$) in PECVD SiO$_{2}$ ($\sim$44°) is approximately 1.8 times larger than in thermal SiO$_{2}$ ($\sim$26°). As described in our track-etching model \cite{dutt_annealing_2023}, the cone angle depends only on the ratio of the track-etch rate to the bulk-etch rate. Hence, the different angles indicate different etch-rate ratios for PECVD and thermal SiO$_{2}$. This discrepancy is not unexpected because PECVD-deposited films typically differ in morphology, density, and stoichiometry compared to thermally grown SiO$_{2}$.

To quantify the bulk etch rates, we measured the thickness etched from each film after etching in 3\% HF for different defined time intervals. The thickness difference before and after etching was measured using ellipsometry which revealed that thermal SiO$_{2}$ etches at 15.6 $\pm$ 0.6 nm/min, while PECVD SiO$_{2}$  etches at 34.3 $\pm$ 1.2 nm/min. Since the cone shape depends only on the ratio of track to bulk etch rate, the significantly different cone angles cannot be explained solely by the variation in bulk etch rates. Therefore, the track etch rate must also differ between the two types of SiO\textsubscript{2}. Using our track-etching model \cite{dutt_annealing_2023} and the measured pore radii, we estimate track etch rates of 69 $\pm$ 3 nm/min for thermal and 90 $\pm$ 6 nm/min for PECVD SiO\textsubscript{2}, respectively.

\begin{figure}
    \centering
    \includegraphics[width=\textwidth]{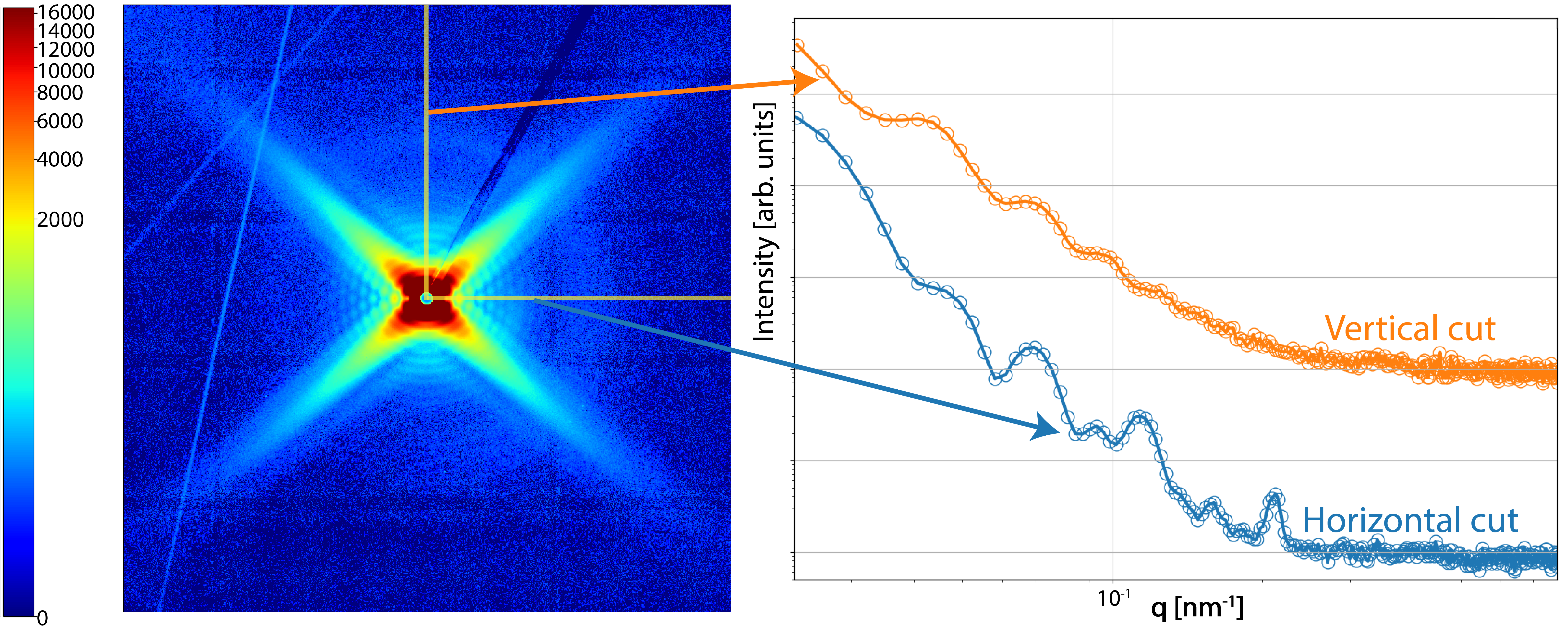}
    \caption{Representative 2D scattering pattern (left) from conical nanopores in thermal SiO$_2$ illustrating the regions used for horizontal (blue) and vertical (orange) cuts. The sample was irradiated with 1.6 GeV Au ions and etched for 15 mins in 3\% HF. Measurements were performed with a tilt angle of the surface mormal of $\sim$20° with respect to the X-ray beam. The corresponding one‐dimensional (1D) intensity profiles (right) are shown as a function of the magnitude of the scattering vector $\vec{q}$.}
    \label{fig:1DSAXS3}
\end{figure}
While the scanning electron microscopy images reveal the variation in nanopore size for PECVD SiO$_{2}$  compared to thermal SiO\textsubscript{2}, as well as differences in nanopore morphology, these images do not provide robust statistical information and are prone to measurement errors. Cross-sectional SEM imaging provides limited statistical reliability, as the probability of cleaving directly through a nanopore's central axis is extremely low. This sampling bias introduces significant uncertainties in dimensional measurements and makes it challenging to obtain robust structural information about the nanopores.

Figure \ref{fig:1DSAXS3} shows a representative 2D scattering pattern obtained from conical nanopores in thermal SiO$_{2}$. This image represents the simultaneous measurement of approximately 10$^7$ parallel nanopores, tilted by $\sim$20° with respect to the X-ray beam. Although fitting the entire image can give precise information on the nanopore size and cone angle, fitting the size distribution is computationally too expensive \cite{hadley_analysis_2020}. Our new approach of fitting the scattering intensities uses two orthogonal 1D cuts of the scattering image (figure \ref{fig:1DSAXS3}). This analysis preserves the high precision of SAXS analysis while substantially reducing computational demands and enabling investigation of the size distributions. \ref{fig:1DSAXS3} highlights the regions selected for horizontal and vertical cuts. The resulting scattering intensity profiles (vertical cut: orange and horizontal cut: blue) from these cuts are shown on figure \ref{fig:1DSAXS3}, right. The intensity values obtained at different tilt angles were fitted as described below.

\begin{figure}
    \centering
    \includegraphics[width=\textwidth]{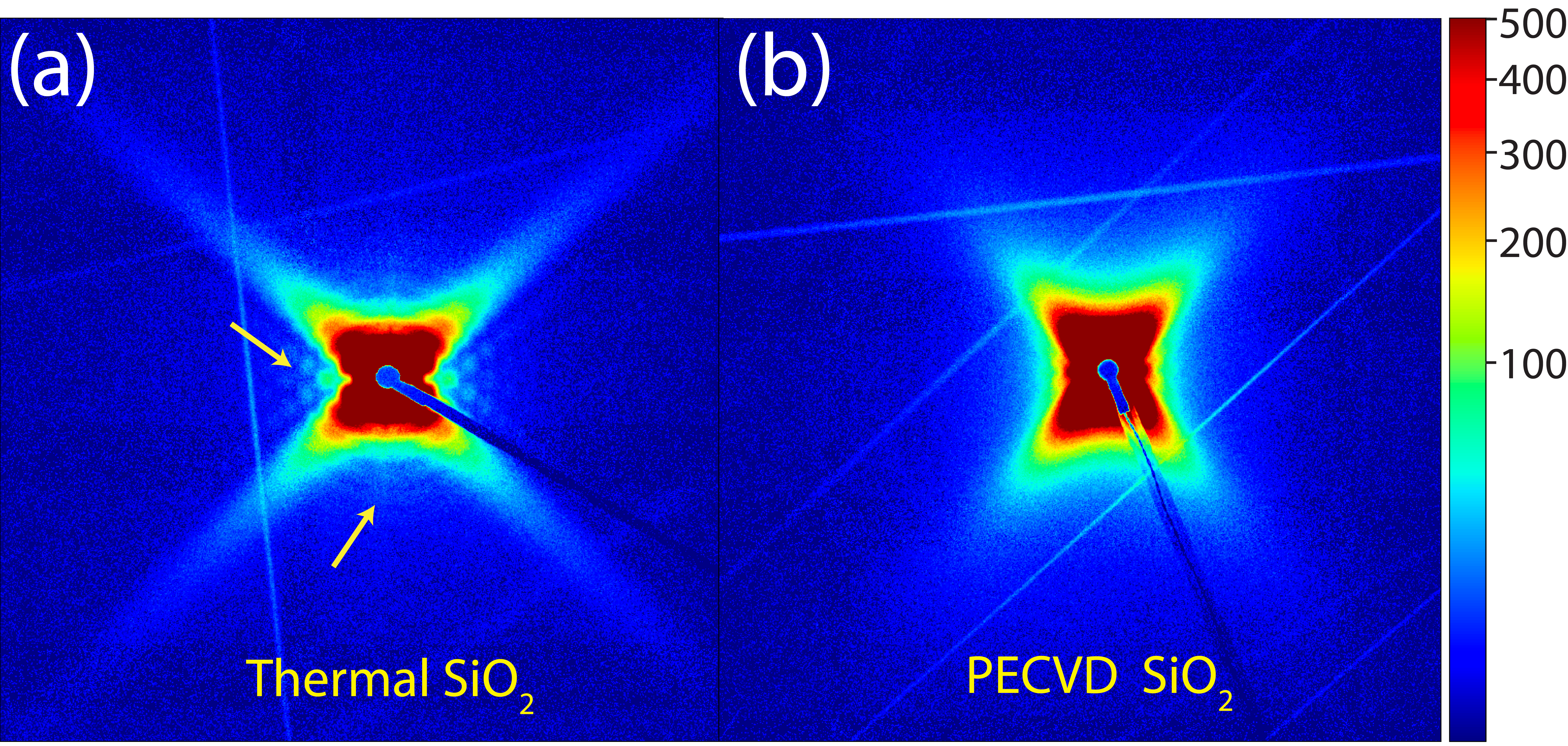}
    \caption{Two‐dimensional small‐angle X‐ray scattering (SAXS) patterns of conical nanopores in thermal (a) and PECVD (b) SiO$_2$ produced from irradiation of thin film samples with 1.6 GeV Au ions and etching with 3\% HF, shown using the same intensity‐contrast scale. The arrows in (a) highlight secondary scattering features that are more pronounced in thermal SiO$_2$ than in the PECVD sample, consistent with a higher polydispersity of pore sizes in the latter.
}
    \label{fig:1DSAXS2}
\end{figure}

Figure~\ref{fig:1DSAXS2} presents two-dimensional scattering images for thermal (a) and PECVD (b) SiO\textsubscript{2}. As indicated by the yellow arrows in (a), clear secondary scattering features can be observed in thermal SiO\textsubscript{2}, indicative for a low dispersity in nanopore dimensions. In contrast, these features are absent in PECVD SiO\textsubscript{2}. We ascribe this effect to the variation in nanopore size, as each pore generates a slightly different scattering intensity, effectively smearing out the secondary features. Furthermore, the absence of secondary features makes it difficult to fit the SAXS data using our 2D fitting model \cite{hadley_analysis_2020}.

\begin{figure}
    \centering
    \includegraphics[width=\textwidth]{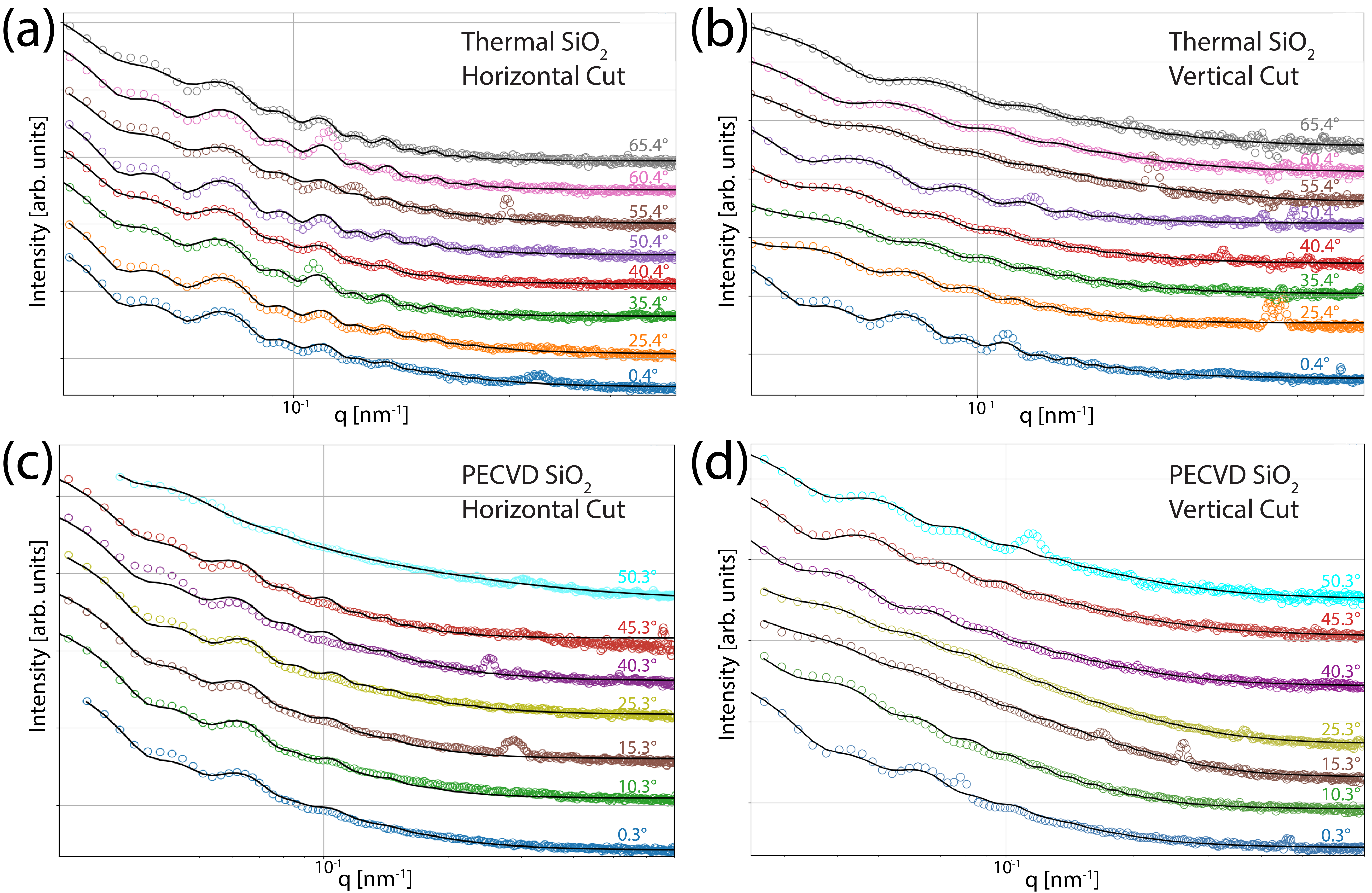}
    \caption{One‐dimensional SAXS profiles of ion‐irradiated thermal (a, b) and PECVD (c, d) SiO$_2$, extracted along the horizontal (a, c) and vertical (b, d) directions. Each pattern corresponds to a different tilt angle (labels in degrees), offset vertically for clarity. The solid lines denote model fits based on conical and core‐transition models.}
    \label{fig:1DSAXS4}
\end{figure}

The scattering intensities from vertical and horizontal cuts of 2D SAXS images were analyzed for both PECVD and thermal SiO$_{2}$ samples using the methodology detailed in the Experimental and Theory section. Figure \ref{fig:1DSAXS4} presents the experimental data and the corresponding model fits, where panels (a,b) represent data from nanopores in thermal SiO$_{2}$ and (c,d) show data from nanopores in PECVD SiO$_{2}$. Horizontal cuts are shown in panels (a,c), while vertical cuts at various tilt angles are shown in panels (b,d). The fitting models demonstrate excellent agreement with the experimental data across all scattering curves. 
Both samples were irradiated with 1.6 GeV Au ions and subsequently etched in 3\% HF. For comparative analysis, we selected samples with similar nanopore radii: thermal SiO$_{2}$ (etched for 12 minutes) yielded pores of average radius 141.3 nm, while PECVD SiO$_{2}$ (etched for 7 minutes) produced pores of average radius 154.2 nm. These PECVD samples exhibited nanopores of high quality, with a size distribution of $\sim$8.3\%. While this size distribution is narrow compared to many nanopore systems \cite{sutariya_realistic_2022}, thermal SiO$_{2}$ nanopores show an even narrower size distribution of only $\sim$2.1\% . The higher dispersity observed in PECVD-based nanopores could be the result of defects or localized variations in material properties. While thermal SiO$_2$ typically exhibits high homogeneity in local material properties, factors such as the etching process and the ion irradiation energy straggling may introduce an effective narrow size distribution of pores. Although we apply a Schulz-Zimm distribution to model the nanopore radius, as described in the Experimental and Theory section, this distribution strongly correlates with variations in the cone angle. We can thus ascribe the polydispersity directly to the variation of the cone angles as well.
The influence of size distribution on the scattering patterns is evident upon detailed examination. The horizontal cut intensities from thermal SiO$_{2}$ nanopores reveal 6-8 distinct oscillations, whereas PECVD SiO$_{2}$ displays a maximum of 4 oscillations. Furthermore, the reduced peak-to-trough amplitude in oscillations resulting from nanopores in PECVD SiO$_{2}$ corroborates the broader size distribution obtained from our fitting analysis. The reduced number of oscillations in PECVD SiO${_2}$ 1D scattering intensity corresponds to  the absence of secondary scattering features in the 2D scattering image as described above.

\begin{center}
    \includegraphics[width=\textwidth]{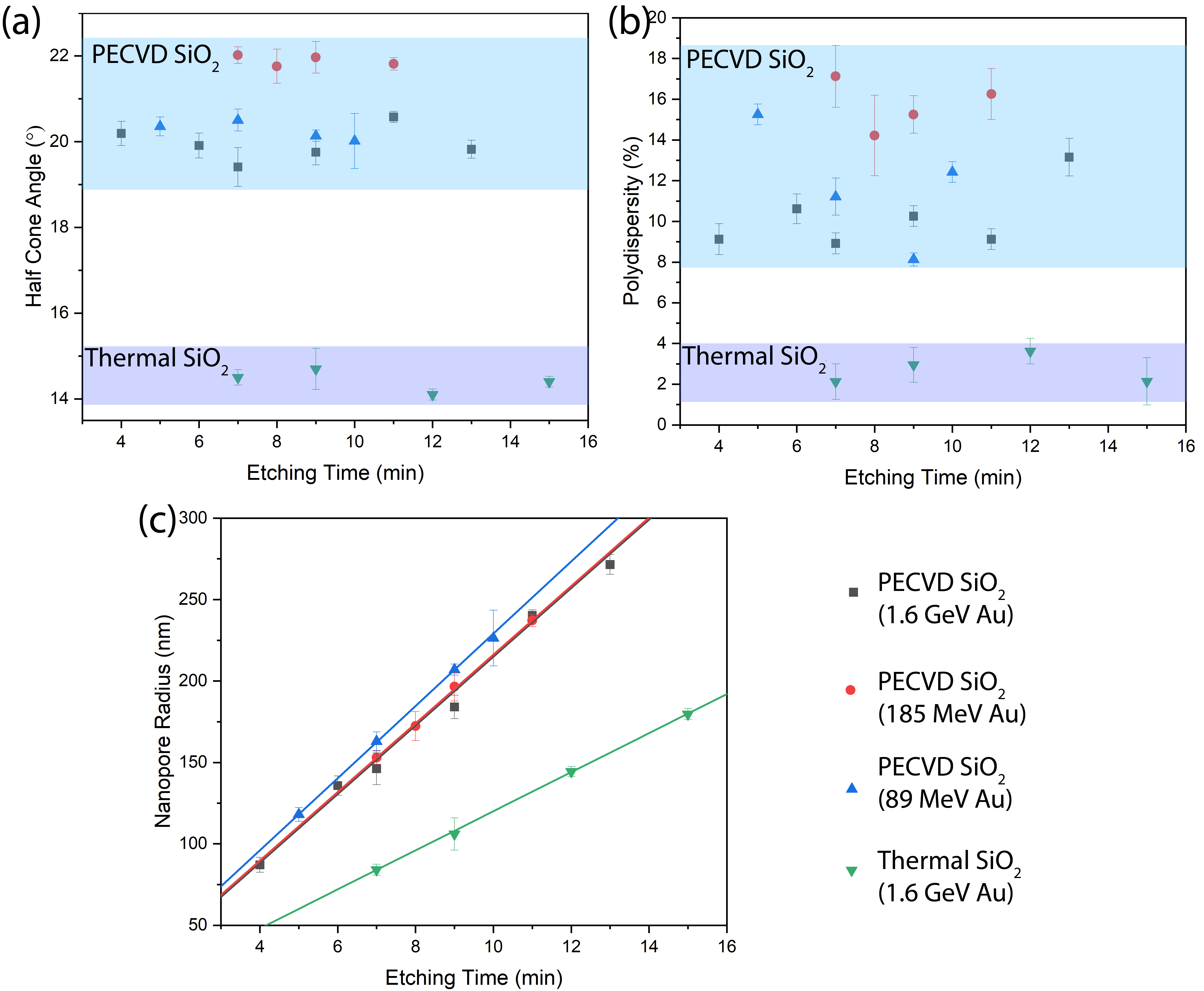}
    \captionof{figure}{%
        Half cone angle (a), percentage polydispersity (b), and nanopore radius (c) as a function of etching time for conical nanopores in thermal and PECVD SiO$_{2}$, irradiated with Au ions at different energies. The shaded regions  in (a) and (b) highlight approximate parameter ranges for thermal versus PECVD SiO$_{2}$. In (c), solid lines represent linear fits to the data. 
    }
    \label{fig:1DSAXS5}
\end{center}

Figure \ref{fig:1DSAXS5} presents the evolution of (a) half cone angle, (b) percentage polydispersity, and (c) nanopore radius as a function of etching time for nanopores in PECVD and thermal SiO$_{2}$ irradiated with Au ions at different energies. The analysis reveals distinct differences between the two types of SiO$_{2}$. Nanopores in PECVD SiO$_{2}$ exhibit on an average $\sim$1.8 times larger cone angles compared to the thermal SiO$_{2}$. Moreover, the size distribution of nanopores, quantified by the polydispersity values (Figure \ref{fig:1DSAXS5}b), are higher ($\sim$8-18\%) in PECVD SiO$_{2}$ compared to ($\sim$2-4\%) in thermal SiO$_{2}$. We note that compared to many other systems, the pore size homogeneity in thermal SiO$_2$ is exceptional.
In PECVD SiO$_{2}$ samples, nanopores fabricated using 185 MeV Au ion irradiation show slightly larger cone angles compared to those created with 89 MeV and 1.6 GeV Au ions. This variation may result from sample-to-sample difference that can arise from the PECVD deposition processes, as the samples originated from different deposition runs. These 185 MeV-fabricated nanopores also exhibited the highest polydispersity, underscoring the variability in PECVD film characteristics.
The validity of our analysis is supported by multiple cross-validation measures. The cone angle values derived from 1D fits for thermal SiO$_{2}$  not only agree well with those obtained from the established 2D fitting model \cite{hadley_analysis_2020} but also correspond well with the cross-sectional SEM images. Furthermore, polydispersity values determined by SAXS correlate strongly with the estimates from scanning electron microscopy analysis. It is important to emphasize that SAXS analysis provides a statistically robust characterization of polydispersity, radius and cone angle values by sampling over $10^7$ nanopores—a population size unattainable through microscopy analysis.
A linear fit of the nanopore radius versus etching time yielded radial etching rates of 21.1 $\pm$ 0.8, 21.1 $\pm$ 0.2, and 22.1 $\pm$ 0.2 nm/min for PECVD SiO${2}$ nanopores fabricated using 1.6 GeV, 185 MeV, and 89 MeV Au ions, respectively. In contrast, thermal SiO${2}$ exhibited a lower radial etching rate of 11.9 $\pm$ 0.1 nm/min. Using these values in conjunction with our track etching model \cite{dutt_annealing_2023}, we calculated track etching rates of 85 $\pm$ 10, 87 $\pm$ 5, and 91 $\pm$ 6 nm/min for the respective PECVD samples, while thermal SiO$_{2}$ showed a track etching rate of 68 $\pm$ 4 nm/min. These track etching rates agrees well with the values calculated using radii from SEM images and employing our track etching model.

\section{Conclusions}

In this study, we performed a comparative analysis of conical nanopores fabricated in thermal and PECVD SiO$_{2}$ using ion track etching employing SEM and SAXS. Our findings reveal substantial differences in the track etching rate and the bulk etching rate between these materials, which in turn affect the nanopore geometry. Nanopores in PECVD SiO$_2$ exhibit cone angles approximately 1.8 times larger than those in thermally grown SiO$_2$—a variation attributable to differences in material density, composition and stoichiometry between the two oxide types. Furthermore, thermal SiO${_2}$ demonstrates remarkable homogeneity (polydispersity $\sim$2-4\%) compared to PECVD SiO${2}$ (polydispersity $\sim$8-18\%). Although PECVD SiO$_{2}$ nanopores show broader size distributions, these values still represent a significant improvement over existing nanoporous systems reported in the literature \cite{sutariya_realistic_2022} as the pore distributions can exceed 50\% in existing systems. The use of different SiO$_2$ compositions allows for tuning of the pore geometry, which can have a significant influence on performance in different applications \cite{ramirez_pore_2008,balannec_nanofiltration_2018,pietschmann_rectification_2013,laohakunakorn_electroosmotic_2015}.
 
The new analytical methodology developed and employed in this study marks a pronounced advancement in conical nanopore characterization. This approach enables reliable assessment of size distributions while maintaining high precision in the determination of nanopore shape, thereby facilitating a detailed investigation of the relationships between fabrication conditions and resultant pore characteristics. The ability to quantify the size distribution with high accuracy is particularly valuable, as size uniformity often plays a crucial role in the performance of nanopore-based applications.

\section{Experimental and Theory}

\subsection{Nanopore formation in Thermal and PECVD SiO$_{2}$}

We utilized two types of amorphous silicon dioxide samples. The first type consisted of 1 $\mu$m thick thermally grown SiO$_{2}$ on <100> Si substrates (300 $\mu$m thickness), obtained commercially from WaferPro Ltd, USA. The second type comprised PECVD-deposited SiO$_{2}$ films ($\sim$1.1 $\mu$m thick) grown on a 300 $\mu$m thick, polished <100> Si substrates using an Oxford Plasmalab 100 PECVD system. PECVD deposition was performed at 650 °C with gas flow rates of 16 sccm SiH$_4$, 980 sccm N$_2$, and 14 sccm NH$_3$. Ellipsometry measurements employing a Tauc-Lorentz model revealed a deposition rate of $\sim$36.6 nm/min.

Both sample types were irradiated with Au ions of 1.6 GeV at the UNILAC accelerator (GSI Helmholtzzentrum für Schwerionenforschung GmbH, Germany). Additionally, the PECVD SiO$_{2}$ samples were irradiated with 89 MeV and 185 MeV Au ions at the 14UD accelerator (Heavy Ion Accelerator Facility, Australian National University). The irradiation fluences ranged from 10$^8$ to 5$\times$10$^8$ ions cm$^{-2}$, ensuring minimal overlap between ion tracks and resulting nanopores \cite{dutt_annealing_2023,riedel_statistical_1979,kluth_measurement_2008}. 

To convert the ion tracks into nanopores, the samples were etched at room temperature in 3\% hydrofluoric acid for varying durations. The etching process was stopped by removing samples from the etchant followed by three successive rinses in de-ionized water, each lasting 30 seconds, after which the samples were air dried. The scanning electron microscopy images of the nanopores were obtained through FEI Verios 460 micrscope. For cross-section images, the samples were cleaved and imaged vertically.

\subsection{Small angle X-ray scattering}
Transmission small angle X-ray scattering (SAXS) measurements were conducted at the SAXS/WAXS beamline at the Australian Synchrotron, Melbourne with a photon energy of 12 keV. The sample-to-detector distances ranged between 7.2 m and 7.6 m. Data collection was performed using Pilatus 1M and Pilatus 2M detectors during different measurement cycles. A silver behenate (AgBeh) standard was used to calibrate both the sample-to-detector distance and beam center positions. Exposure times ranged from 2 to 10 seconds, with samples mounted on a three-axis goniometer for precise alignment with the incident X-ray beam. Detailed information regarding alignment, tilts, measurements, geometry, and 2D analysis procedures can be found in our previous works \cite{hadley_etched_2019,hadley_analysis_2020}. 

The two-dimensional scattering patterns were converted into one-dimensional scattering intensities through horizontal and vertical cuts (along q$_x$ and q$_y$ respectively) originating from the center of the beamstop. We selected the cut with minimal interference from the Kossel line by comparing the symmetric positive and negative values q$_x$ and q$_y$ relative to the center of the beam. These cuts were obtained through azimuthal integration along the masked region (see Fig. \ref{fig:1DSAXS3}). To preserve the accuracy of polydispersity measurements without averaging interference effects, the cuts were kept as narrow as practicable.
For analysis of the vertical cut, we employed our previously reported cone model \cite{hadley_analysis_2020}, where the form factor is given by:
\begin{equation}
\label{eq1}
f_{2D}(q_r, q_z) = C \int_{0}^{L} J_{1}\bigl(r_z,q_r\bigr),\frac{r_z}{q_r},\exp\bigl(-\mathrm{i},z,q_z\bigr),\mathrm{d}z
\end{equation}
Here, $f_{2D}(q_r,q_z)$ represents the form factor assuming rotational symmetry along the Z-axis of the conical nanopores, $L$ is the length of the conical nanopores, $J_1$ denotes the first-order Bessel function, and C accounts for electron density contrast and other constant parameters. The radial component of the scattering vector ($\vec{q}$), denoted as $q_r$, is given by $\sqrt{q_x^2 + q_y^2}$. For the vertical cut analysis, we set $q_x = 0$, which reduces $q_r$ to $q_y$. This formulation captures the scattering amplitude for conical objects while accounting for the radius variation along the Z-axis.
The horizontal cut analysis was performed setting $q_y = 0$. Equation \ref{eq1} then reduces to a 'core transition' model, previously detailed in our work \cite{wang_saxs_2022,dutt_role_2023}. This model incorporates a constant core radius (fixed to the ion track radius determined by SAXS \cite{kluth_fine_2008,mota-santiago_nanoscale_2018,dutt_annealing_2023}) with a linear density transition region. The corresponding form factor is expressed as:
\begin{equation}
\begin{aligned}
f_{CT}(q_r, q_z)
&= 4\pi \,\frac{\sin\!\bigl(q_{z} \,l\bigr)}{q_{z}} \,\frac{\pi}{(q_{r})^{2}}
\\[6pt]
&\quad\times \Bigl[
      -\,R_C\,J_{1}\bigl(R_C\,q_{r}\bigr)\,H_{0}\bigl(R_C\,q_{r}\bigr)
      + (R_C + R_T)\,J_{1}\bigl((R_C + R_T)\,q_{r}\bigr)\,H_{0}\bigl((R_C + R_T)\,q_{r}\bigr)
\\[4pt]
&\qquad\quad
      +\,R_C\,J_{0}\bigl(R_C\,q_{r}\bigr)\,H_{1}\bigl(R_C\,q_{r}\bigr)
      - (R_C + R_T)\,J_{0}\bigl((R_C + R_T)\,q_{r}\bigr)\,H_{1}\bigl((R_C + R_T)\,q_{r}\bigr)
   \Bigr]
\end{aligned}
\end{equation}

where $R_c$ represents the fixed core radius matching the ion track radius, $R_T$ denotes the transition region thickness, and $H_u(z)$ represents the Struve function given by:

\begin{equation}
H_u(z) 
= \left(\frac{z}{2}\right)^{u+1}
  \sum_{m=0}^\infty 
    \frac{(-1)^m \left(\frac{z}{2}\right)^{2m}}
         {\Gamma\!\bigl(m + \tfrac{3}{2}\bigr)\,\Gamma\!\bigl(m + u + \tfrac{3}{2}\bigr)}
\end{equation}

 To measure the distribution of the nanopore sizes, we implemented a narrow Schulz-Zimm distribution \cite{wang_saxs_2022,dutt_annealing_2023,mota-santiago_nanoscale_2018,dutt_role_2023}. Readers are referred to these works for detailed information on the implementation of polydispersity. The fits are performed using a custom C- and Python-based code that employs a non-linear least-squares algorithm. To correct for background scattering originating from the air, the substrate, etc., we employ a q-dependent background \cite{kiy_ion_2021} as it is not feasible to extract and subtract the background explicitly for these materials systems. 

We fit the horizontal cuts to determine the nanopore radii and the corresponding size distribution. The oscillations in the scattering intensity along the horizontal cuts remain the same as a function of the tilt angle between cone axis and X-ray beam, making it impossible to extract direct information about the cone angles. In contrast, the scattering intensities in the vertical cuts vary with the tilt angle and provide information on the cone angle of the nanopores. As the cone angle and tilt angle are highly correlated, fitting just a single vertical cut produces large uncertainties mainly due to the challenges in accurately determining the experimental tilt angle \cite{hadley_analysis_2020,hadley_etched_2019}. Therefore, we acquire scattering images at multiple tilt angles and fit the resulting scattering intensities simultaneously.

Our overall fitting strategy proceeds as follows. For each sample, first, we fit the scattering intensities from individual horizontal cuts obtained at different tilt angles to obtain initial estimates of nanopore radii. Next, multiple horizontal cuts originating from different tilt angles are fitted together to refine the radius values and determine size dispersity more accurately. We then used these refined values as starting points for simultaneously fitting multiple vertical cuts, treating the tilt angle as a variable to account for the imperfect alignment of the cones with the incoming X-rays. It should be noted that the difference in different tilt angles is fixed and known for different experiments. Finally, we combine both horizontal and vertical cuts in a single simultaneous fit, constraining the radius and the cone angle to the same values.

\section{CRediT author statement}
\textbf{Shankar Dutt:} Conceptualization, Methodology, Investigation, Software, Formal analysis, Visualization, Writing - Original Draft.

\textbf{Rudradeep Chakraborty:} Investigation, Formal analysis.

\textbf{Christian Notthoff:} Methodology, Investigation, Software, Writing - Review \& Editing.

\textbf{Pablo Mota-Santiago:} Investigation, Writing - Review \& Editing.

\textbf{Christina Trautmann:} Investigation, Writing - Review \& Editing.

\textbf{Patrick Kluth:} Conceptualization, Methodology, Investigation, Visualization, Writing - Review \& Editing, Supervision, Funding acquisition.

\begin{acknowledgements}
Part of the research was undertaken at the SAXS/WAXS beamline at the Australian Synchrotron, part of ANSTO, and we thank the beamline scientists for their technical assistance. We also acknowledge access to the NCRIS (National Collaborative Research Infrastructure Strategy) supported Heavy-Ion Accelerator facility (the 14UD accelerator) at the Australian National University. The results presented here are based on a UMAT experiment that was carried out on the UNILAC X0 beamline at the GSI Helmholtz Center for Heavy Ion Research, Darmstadt (Germany) in the frame of FAIR Phase-0. This work used the ACT node of the NCRIS-enabled Australian National Fabrication Facility (ANFF-ACT). The authors also acknowledge financial support from the Australian Research Council (ARC) under the ARC Discovery Project Scheme.
\end{acknowledgements}

\bibliography{references}

\end{document}